\documentclass{aa}
\usepackage{txfonts}
\input epsf.sty
\input psfig.sty
\def\l({\left(}
\def\r){\right)}

\def\be{\begin{equation}}
\def\ee{\end{equation}}

\def\th{\vartheta}
\def\ph{\varphi}
\def\pt{\partial}

\def\omg{\Omega}

\def\d{\rm{d}}

\def\df{\widehat{=}}
\def\mn{\mu\nu}

\begin{document}

\title{Evolutionary sequences of rotating protoneutron stars}
\author{Lo\"{\i}c Villain \inst{1,2,3} \and Jos\'e A. Pons \inst{1,4}
\and Pablo Cerd\'a--Dur\'an \inst{1} \and Eric Gourgoulhon \inst{3}}
\institute{ Departament d'Astronomia i Astrof\'{\i}sica, Universitat
de Val\`encia, E-46100 Burjassot, Spain,
\and Copernicus Astronomical Center (CAMK), Polish Academy of Sciences,
Bartycka 18, PL-00-716 Warszawa, Poland,
\and Laboratoire de l'Univers et de ses Th\'eories
(UMR 8102 du C.N.R.S.), Observatoire de Paris -- \\
Section de Meudon, F-92195 Meudon Cedex, France,
\and Departament de F\'{\i}sica Aplicada, Universitat d'Alacant,
Apartat de correus 99, 03080 Alacant, Spain.}

\offprints{L. Villain, \email{loic@camk.edu.pl}}
\date{Received...../ Accepted.....}

\abstract{
We investigate the evolution of rigidly and differentially rotating 
protoneutron stars during the first twenty seconds of their life. 
We solve the equations describing stationary axisymmetric configurations 
in general relativity coupled to
a finite temperature, relativistic equation of state, to obtain a sequence
of quasi-equilibrium configurations describing the evolution of newly
born neutron stars. 
The initial rotation profiles have been taken to mimic the situation
found immediately following the gravitational collapse of rotating
stellar cores. By analyzing the output of several models, we estimate that
the scale of variation of the angular velocity in a newly born neutron star
is of the order of 7-10 km. We obtain the maximum rotation frequency
that can be reached as the protoneutron stars
deleptonizes and cools down, as well as other relevant parameters such as
total angular momentum or the instability parameter $|T/W|$.
Our study shows that imposing physical constraints (conservation of baryonic mass
and angular momentum) and choosing reasonable
thermodynamical profiles as the star evolves gives results
consistent with the energetics of more complex simulations of non-rotating
protoneutron stars. It appears to be unlikely that 
newly born protoneutron stars formed in nearly axisymmetric core collapses
reach the critical angular velocity to undergo the bar mode instability.
They could, however, undergo secular or low $|T/W|$ rotational instabilities
a few seconds after birth, resulting in a strong emission of gravitational
waves retarded with respect to the neutrino luminosity peak.
We also found that the geometry of strongly differentially rotating protoneutron stars
can become toroidal-like for large values of the angular velocity, before reaching
the mass shedding limit.
}

\authorrunning{L. Villain et al.}
\titlerunning{Evolutionary sequences of rotating PNSs}

\maketitle

\section{Introduction}

The first minute of life of non-rotating protoneutron stars (PNS) has 
been studied in detail since the mid 80s by several authors 
(Burrows \& Lattimer \cite{bl86};
Keil \& Janka \cite{kj95}; Pons et al. \cite{pns99,pns00,pns01}). 
The initial hot, lepton rich remnant, left behind following a successful core
collapse supernova is known to evolve in a timescale of tens of seconds to form a cold
\mbox{($T<10^{10}$ K)} catalyzed neutron star. This process is usually 
classified
in three main stages: i) the {\it mantle contraction} during which fast cooling of the outer
regions takes place in about \mbox{0.5 s}, with probably significant accretion;
ii) {\it deleptonization} and consequently heating of the internal core as
energetic neutrinos diffuse out leaving most of their chemical energy 
(analogously to Joule effect) on the way out;
and iii) {\it cooling} by means of diffusion of (mostly) thermal neutrinos, resulting
in a decrease of temperature from about 40-50 MeV to below 2-4 MeV, point at which
the star becomes transparent to neutrinos.
 
Less is known about the early evolution of PNSs that are rapidly rotating, because a fully
consistent study is a formidable task. It requires solving the neutrino transport
equations in at least two-dimensions, coupled to general relativistic hydrodynamics
that describes the fluid motion in rotating relativistic stars.
In addition, the microphysical inputs also need to be treated carefully. As has been
shown previously (Pons et al. \cite{pns99}), internal consistency between 
the equation of state of
dense matter and the neutrino opacities describing the interaction between neutrinos
and matter is needed to achieve reliable results.

A few attempts to treat the problem in a simplified way exist 
(Romero et al. \cite{rom92}; Goussard et al. \cite{G97,G98}; Sumiyoshi et al. \cite{sumi99}; Strobel et al. \cite{SSW99};
Yuan \& Heyl \cite{YH03})
based on the analysis of temporal sequences of
quasi-equilibrium models of PNS, imposing some ad-hoc thermodynamical
profiles ({\it i.e.}, constant temperature, constant entropy and/or 
constant neutrino fraction).
The assumption of quasi-equilibrium is well justified, since the hydrodynamical
timescale ($10^{-3}$ s) is much smaller than the timescale in which 
substantial thermodynamical changes occur (diffusion timescale $\approx 1$ 
s).
However, it is unclear whether or not an isothermal or isentropic profile
can effectively mimic reality. In addition,
there are some physical constraints such as conservation of energy (the total
gravitational mass must  decrease consistently with the neutrino luminosity), 
or conservation of angular momentum, that need to be satisfied when one 
studies a temporal sequence of
quasi-equilibrium models of PNS. These constraints  help to reduce the
parameter space as one calculates the thermodynamical evolution, although
there is much uncertainty about the transport of angular momentum. In general
angular momentum losses by neutrinos are supposed to be small 
(Kazana \cite{Kaz77}).
Neutrinos could redistribute or take away a fraction of the initial angular 
momentum, but quantifying this amount needs of multidimensional
transport simulations. It is equally controversial how important is turbulent
transport, or the role of gravitational radiation. Because of these many
unresolved issues it is usually assumed that the angular
momentum is approximately conserved during the Kelvin-Helmholtz stage.
In addition, the presence of large magnetic field could be crucial. 
It has been recently shown that turbulent mean--field dynamo action
can be effective for PNSs with periods shorter than 1 s, which would
generate very strong magnetic fields in the interior (Bonanno et al.
\cite{BRU}).

The whole problem is far to be solved, and will require some serious 
computational effort but, in the meantime,
the intention of this paper is to begin to understand qualitatively what kind of outcome
will be revealed by such numerical simulations, prior to engaging large scale simulations.
In this line, we improve in several respects the few existing previous works.
First, instead of isentropic or isothermal models, we use realistic profiles coming 
from 1D simulations, conveniently rescaling the
temperature and chemical profiles as functions of density; second, we do not restrict
ourselves to the rigid rotation case, as most of previous studies, 
and include differential rotation in our analysis;
last but not least, we check that fixing the baryonic mass and the
angular momentum results in  some reasonable time dependence to the integrated
relevant physical quantities such as the gravitational mass.
In this way, we can calculate a complete evolutionary sequence of a PNS . 

The plan of the paper is as follows. In \ref{sec:rela} we briefly review the relativistic
description of rotating stars, and the numerical method used to obtain equilibrium
configurations. A similar brief description of the thermodynamics and the equation
of state is presented in \ref{sec:ther}, since for both issues the details have been published
elsewhere. Section \ref{sec:inimod} discusses the choice of parameters for the initial models.
Our results are presented in section \ref{sec:resnum}.
In the first part we describe rigidly rotating PNSs, while the
description of differentially rotating PNSs is given in \ref{sec:difrot}. In both cases, 
we give all the relevant parameters for maximally rotating configurations as well
as the evolution of more realistic sequences. Finally, our main conclusions and
findings are summarized and discussed in \ref{sec:conc}.

\section{Numerical code for stationary axisymmetric spacetime} \label{sec:rela}

 Since the time scale for the matter and the 
gravitational field to settle in stationary equilibrium is much
shorter than all other involved times\footnote{unless a gravitational
dynamical instability occurs.}, a simplifying but plausible
approximation consists of mimicking the time evolution of the spacetime by
obtaining a collection of stationary, axisymmetric and asymptotically
flat snapshots, all of them being solutions of the Einstein equations,
but with a source term that depends on the snapshot.\\

  This approximation has already been adopted by several authors in 
the past (Goussard et al. \cite{G97,G98}; Strobel et al.
\cite{SSW99}; Yuan \& Heyl \cite{YH03}), our code being moreover very
similar to the one of Goussard et al. (\cite{G97,G98}), since we
also employ a fully relativistic spectral code based on the work of
Bonazzola et al. (\cite{BGSM}).  In the following, this article
will be referred to as BGSM, and we address the interested reader to it for more
details about the algorithms. Let us just remind that the BGSM approach
uses the $3+1$ formulation of general relativity, maximal
slicing-quasi-isotropic coordinates, with the assumption that matter
is free of convective motions.  The latter hypothesis is probably too
strong, since  PNSs are known to be convectively unstable
(Keil et al. \cite{KJM96}; Miralles et al.
\cite{MPU}),  but it implies the circularity of the stationary axisymmetric
spacetime (Carter \cite{c69}) and greatly simplifies
the calculations. \\ 

Under these approximations, the metric is written
\footnote{For all this article, we are using natural units
in which the speed of light is equal to unity.}
\begin{eqnarray} \label{e:met_msqi}
{\d} s^2 \, &\df& \, g_{\mn}\, {\d} x^{\mu} {\d} x^{\nu}\,
\nonumber \\
&\df& \, - \l(N^2\,-\,N_{\ph}\,{N^{\ph}}\r) {\d t}^2 - 2\,N_{\ph}\, {\d t}
\,{\d\ph}\,\nonumber\\
& +&\,\frac{A^4}{B^2}\l({\d r}^2\,+\,r^2\,{\d\th}^2\,+\,B^4\,r^2\,
\sin^2 \th\,{\d\ph}^2\r)\,,
\end{eqnarray}
where we have introduced the usual notation of the $3+1$ formalism:
$N$ is the lapse, $N^{\ph}$ is the third component of the shift
$3$-vector, with \mbox{$N_{\ph}\,\df\,g_{\ph i}\,N^i\,=\,A^4\,B^2\,
N^{\ph}r^2\,\sin^2 \th$} being the covariant $\ph$-component of the
latter. Thanks to the hypothesis of axisymmetry and stationarity, the functions 
$N, N^{\ph}, A$ and $B$ depend only on $r$ and $\th$ .\\

Despite the general formalism and structure of the code is
very similar to that of  Goussard et al. (\cite{G97,G98}), 
there are a few differences that should be mentioned:
\begin{itemize}
\item[-] we use a more recent version of the spectral BGSM code (see
Gourgoulhon et al. \cite{GHLPBM}), in which at least two shells
map the interior of the star, in order to better describe the
non-spherical surface. This code is based on the C++ library LORENE\footnote{
{\rm http://www.lorene.obspm.fr}}, a software package for numerical relativity
freely available under GNU license;
\item[-] the equation of state (EOS) is written in tabular form and 
interpolated using cubic Hermite polynomials, as defined 
in Nozawa et al. (\cite{NSGE}),
according to a general technique presented in Swesty (\cite{S96}) that 
preserves thermodynamical consistency (see also further remarks below);
\item[-] during the evolution of PNSs we do not restrict
ourselves to 3 or 4 typical evolutionary snapshots that are
just isentropic or isothermal profiles. Instead we adopt more
realistic profiles, rescaled from the results of 1D simulations
with neutrino transport (Pons et al.
\cite{pns99}, more details are given in the next section). 
\end{itemize}

For clarity, let us be more specific about the last remark. From the
hydrodynamical point of view, one must deal with the conditions for
the existence of stationary solutions of the Einstein equations in
presence of a rotating perfect fluid (see also BGSM and Goussard et
al.). The main issue is that the relativistic Euler equations, derived
from the conservation of the energy-momentum tensor projected into a
space-like $3$-surface, are reduced in the case of stationary rotating
motion to
\be \label{e:stmo}
\frac{\pt_i\,P}{e\,+\,P}\,+\,\pt_i\,\ln
\left[\frac{N}{\Gamma}\right]\, =\, -\,F\,\pt_i\,\omg\,, \ee where
\begin{itemize}
\item[-] $P$ is the pressure and $e$ the total energy density, both
measured in the comoving frame;
\item[-] $N$ is the already defined lapse function, $\Gamma$ the
Lorentz factor linking the Eulerian observer and the comoving
observer, $\Omega\,\df\,{u^\ph}/{u^t}$ the angular velocity, and $F\,\df\,u_\ph\,u^t$
with $u$ being the $4$-velocity of the fluid;
\item[-] the index $i$ is either $1$ or $2$, since all the previous
quantities depend only on $r$ and $\th$.
\end{itemize}

 As explained in BGSM  Eq. (\ref{e:stmo})
does not admit a first integral for arbitrary thermodynamics or rotation
profiles \mbox{$\omg[r,\th]$}, but it does in some simple cases. 

Concerning 
rotation, the right hand side of equation (\ref{e:stmo}) is a first
integral if either $\omg$ is constant in space
(rigid rotation), or the function $F$ can be locally written as
\mbox{$F\,=\,F[\omg]$}. The latter leads to the relation
\be \label{e:relf}
F[\omg]\,-\,\frac{A^4\,B^2\,r^2\,\sin^{2}\th\,(\omg\,-\,N^\ph)}
{N^2\,-\,A^4\,B^2\,r^2\,\sin^{2}\th\,(\omg\,-\,N^\ph)^2}\,=\,0\,,
\ee
between $\omg$ and the coefficients of the metric, which allows the
determination of the profile of angular velocity.

For the differential rotation, we adopt the law proposed by Komatsu et al.
(\cite{keh89}) and used by many authors 
(Komatsu et al. \cite{keh89};
Goussard et al. \cite{G97,G98}; Baumgarte et al. \cite{Bau00})
\be \label{e:lawdif}
F[\omg]\,=\,R_{0}^2\,(\omg_c\,-\,\omg)\,,
\ee
where $\omg_c$ is the limit of the function $\omg[r,\th]$ on the rotation axis
and $R_0^2$ is a parameter with the dimension of the square of a 
length, noted $A$ in Komatsu et al. (\cite{keh89}). We shall 
discuss later our
choices for the values of these parameters. It must be remarked that
despite its simplicity, this law gives reasonable rotation profiles qualitatively
similar to those obtained from core collapse simulations (see section \ref{sec:inimod}).

 Regarding the EOS, 
it can be shown that isentropic or isothermal profiles are sufficient 
conditions for Eq. (\ref{e:stmo}) to be integrable. Yet another possibility 
which is less restrictive exists : assuming that the total energy density $e$
can be written as an {\it effective} function of the pressure
$e\,=\,e[P]$. Indeed, in this case, even if the actual EOS is not barotropic, assuming
that the EOS is a one--parameter function ensures
the existence of the first integral
\be \label{e:intpsenth}
\widehat{H}\,+\,\ln\left[\frac{N}{\Gamma}\right]\,
+\,\int{F[\omg]\,{\d}\omg}\,=\,\textrm{const.}\,, 
\ee
where the function $\widehat{H}$ is defined by
\be \label{e:dfpsenth}
\widehat{H}\,\df\,\int{\frac{{\d} P}{e[P]\,+\,P}}\,,
\ee
with suitable boundary conditions. In this work, we have rescaled
the temperature and composition profiles assuming, at each evolutionary
time, that they depend only on the baryon density $n_B$.
In this way, the EOS behaves {\it effectively} as a barotrope, and
the pseudo-enthalpy function (\ref{e:dfpsenth}) can be written as
\be \label{e:dfpsenth2}
\widehat{H}\,\df\,\int_0^{n_B} {\frac{{\d} P[n_B]}{{\d} n_B}\,
\frac{1}{e[n_B]\,+\,P[n_B]}{\d} n_B}\,,
\ee
with the effective functions $P[n_B]$ and $e[n_B]$ depending
only on the baryonic density $n_B$.

At zero temperature, $\widehat{H}$ is equal to the logarithm of the
specific enthalpy
\be \label{eq:enth.t.nulle}
H\,=\,\ln\left[\frac{e\,+\,P}{n_B\,m_B}\right]\,,
\ee
where $m_B$ is the baryon mass. 
As a basic test of the numerical method of integration, we have checked, 
in the case of tabulated EOS at \mbox{$T\,=\,0$}, the agreement between 
$\widehat{H}$ calculated numerically and the analytical 
result (\ref{eq:enth.t.nulle}).  Due to the finite number of points 
in the tabular EOS the agreement is not complete, but the discrepancy was always 
found to be smaller than $1\%$ in all tested cases (even
with strong differential rotation).

\section{Equation of state  and thermodynamics} \label{sec:ther}

Most of previous studies of structure of rotating neutron stars
have only been performed with polytropes, or at most zero
temperature EOS. This is not enough for the purpose 
of studying PNS evolution, since thermal and chemical effects are 
relevant and need a careful treatment.
Only in a few works 
(Goussard et al. \cite{G97,G98}; Strobel et al. \cite{SSW99})
a realistic, finite temperature EOS has been used. The works by Goussard et al.
(\cite{G97,G98}) used a Skyrme-like potential model (Lattimer \& Swesty \cite{LS91}), while
the more recent work by Strobel et al. (\cite{SSW99}) used a Thomas-Fermi model
with momentum and density dependent effective nucleon-nucleon interaction
developed by Myers \& Swiatecky (\cite{MS90}). 

In this paper, we employ an EOS in the framework of relativistic field
theory, in which the interaction between nucleons is mediated by 
the exchange of three meson fields. The coupling constants are chosen
to reproduce some bulk properties of nuclear matter such as the nuclear
saturation density, symmetry energy or nuclear compressibility.
The formalism is widely used, and to avoid unnecessary repetition we refer
to the complete review by Prakash et al. (\cite{Pra97}) and references therein.
The particular choice of the coupling constants is detailed in Pons et al. 
(\cite{pns99}).

\begin{figure}
\psfig{figure=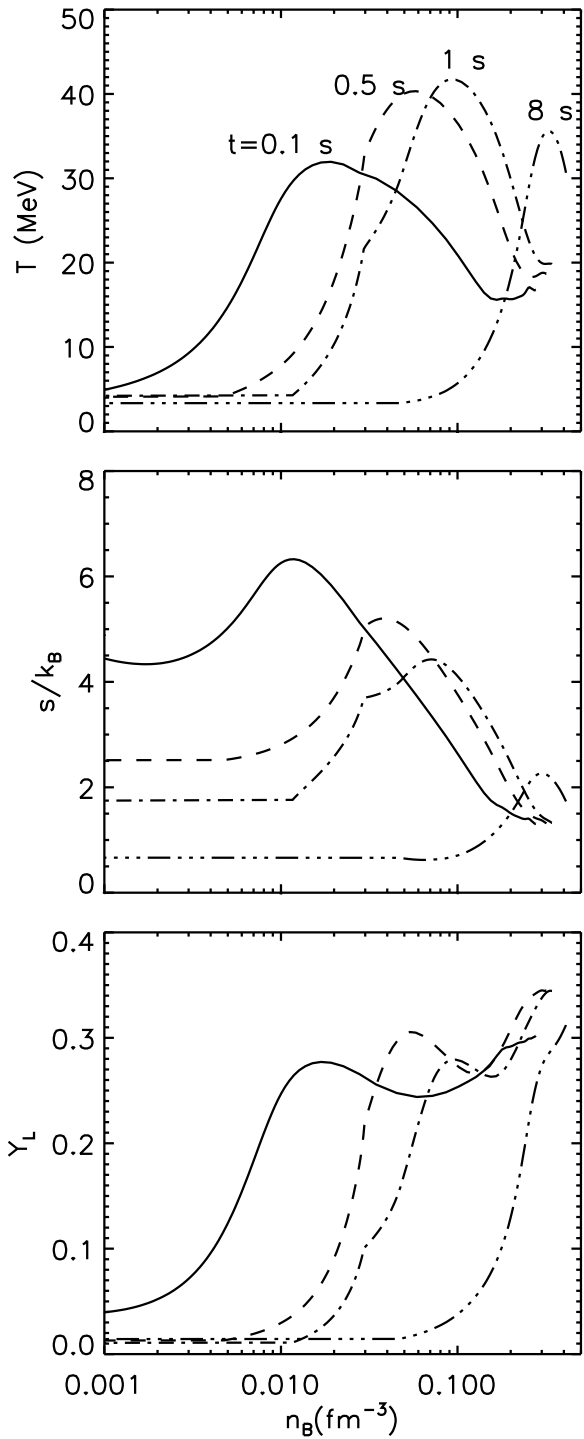}
\caption{Temporal variations of the temperature (top), entropy (middle) 
and lepton fraction (bottom) as functions of the baryon density, for 
evolutionary times corresponding to $t=0.1, 0.5, 1$ and 8 seconds after 
formation of the PNS.
Results are for a $M_B=1.6 M_\odot$ star in spherical symmetry.
}
\label{fig:eos}
\end{figure}

Figure \ref{fig:eos} shows profiles from 1D simulations of a 
\mbox{$M_B=1.6~ M_\odot$} 
spherical star from Pons et al. (\cite{pns99}). In previous works
(Goussard et al. \cite{G97,G98}; Strobel et al. \cite{SSW99})
the evolution of the star was approximated by choosing either simplified profiles
of constant entropy or temperature, or the combination of a constant
low entropy core ($s=1-2$) surrounded by a high entropy mantle ($s=4-5$) to
mimic the situation in the early Kelvin-Helmholtz stage. It was shown
that the choice of the critical density that separates the hot envelope
from the inner core affected substantially the results. 
To reduce the uncertainties associated to the choice of this critical
density and the fact that PNSs are not isothermal or isentropic configurations,
we have chosen to use the thermodynamical structure from 1D simulations as a 
better approximation to reality, and assume that at each time the temperature 
and lepton fraction have the same dependence on the baryon density as 
the non-rotating models, $T=T[n_B]$ and $Y_L=Y_L[n_B]$.
More explicitly: given the results from 1D simulations, the profiles of density,
temperature, lepton fraction, entropy and pressure are tabulated at each time step, 
and these tables are taken as effective one--parameter EOSs where the independent
variable is the baryon density ($n_B$) and the rest of variables are functions of 
only $n_B$.

Notice that, as we explained in the previous section, this (pseudo--barotropicity) 
assumption is a sufficient condition to ensure that the equation of motion
has a first integral, making affordable the computation of the models.
Of course, to know the exact profiles consistently, one should solve the neutrino 
transport equations in 2D along the whole evolution, but this requires a 
large technical and computational
effort that is out of the scope of this paper.

\section{Choice of the initial models} \label{sec:inimod}

One of the first issues that must be addressed to study the early evolution of neutron
stars is the initial model. In Pons et al. (\cite{pns99}) it was shown that for
spherical models the overall evolution is not very sensitive to modifications of the
initial thermodynamical profiles, while the total mass of the star is the parameter 
that mostly affects the subsequent evolution. It seems therefore natural to chose a canonical
star with a baryon mass 1.6 $M_\odot$,
that corresponds to a gravitational mass of the old, cold configuration
of about 1.44 $M_\odot$. For completeness, we have also studied a model with a
baryonic mass of 1.2 $M_\odot$.
Notice that one cannot use the effective barotropes obtained for a PNSs with a
given mass in the calculations of sequences with a different mass, because
the thermodynamic properties are different.

\begin{figure}
\psfig{figure=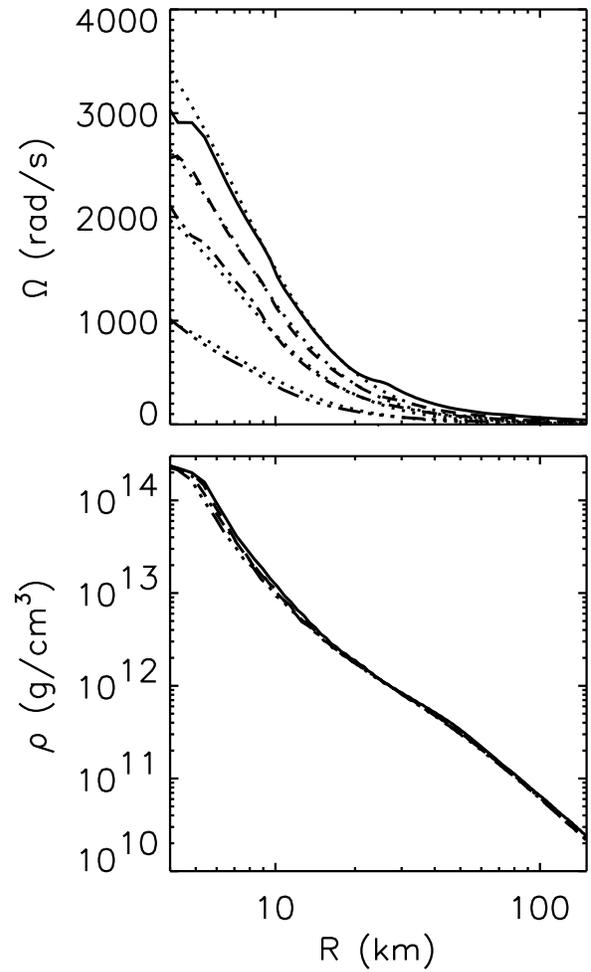}
\caption{Equatorial profiles of angular velocity (top), and density (bottom) 
of PNSs from axisymmetric simulations of stellar core collapse (DFM).
The four models correspond to a different amount of angular momentum of the
iron core, namely, $|T/W|=$ 0.9\% (solid), 0.5\% (dashes), 0.25\% 
(dash-dot),
and 0.05\% (dash-3 dots). The dotted lines on the upper panel are fits to the
simple law (\ref{omelaw}) with $R_0^2=50$ km$^2$.
}
\label{fig:modin}
\end{figure}

\begin{figure}
\psfig{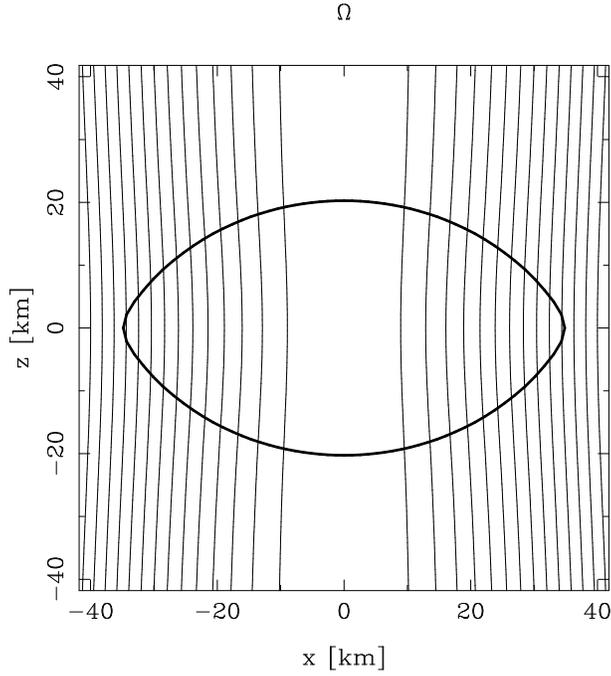}
\caption{Surfaces of constant angular velocity (projected in a transversal plane) 
of a differentially rotating star with $R_0=100$ km, for \mbox{$t=1$ s}, 
and rotating with $\Omega = 354$ rad/s. ($\Omega/\Omega_K=0.995$). The surface
is marked by the thick solid line. Because of relativistic effects, the stationary
solution does not show exact cylindrical symmetry.
\label{fig:ome1}
}
\end{figure}

More difficult is to guess the particular rotation properties of a newly born
neutron star. 
Newly born neutron stars are expected to rotate non rigidly, as opposite
to old neutron stars in which the various viscous mechanisms had enough
time to act. Even in the case that the iron core of the Supernova progenitor was rigidly
rotating (a quite reasonable assumption, since it is an old massive star), the
process of collapse would generate a significant amount of differential rotation.
Indeed, recent calculations of
stellar evolution indicate that the iron core of massive stars is rotating
almost rigidly (Heger et al. \cite{Heg00, Heg03})  
and this is the approximation used in the most recent 
simulations of rotating stellar core collapse (M\"uller et al. \cite{Ewa03}).
Based on this works, we have decided to study a few models from Dimmelmeier et al.
(\cite{DFM}, DFM hereafter) and to run some more simulations of similar models
of stellar core collapse varying the total initial angular momentum. Our models
correspond to those labeled by A1 in DFM, {\it i.e.}, the iron core of the progenitor
is almost rigidly rotating. 
In Figure \ref{fig:modin}, we show profiles of the angular velocity (upper panel) and density
(lower panel) of the inner 150 km, as a function of the equatorial radius,
immediately after the PNS is formed. The different
lines correspond to models with a different amount of initial angular momentum,
with values of the ratio between rotational to gravitational potential energy
($|T/W|$) of 0.9\% (models B3 in DFM), 0.5\% (B2), 0.25\% (B1) and 0.05\%, respectively.
We included the last lower value because it is more realistic (Heger et al. \cite{Heg03};
M\"uller et al. \cite{Ewa03}) than
those studied in DFM. In the figure we also show with dots the simple fit
\be
\label{omelaw}
\omg[r,\th]\,=\omg[r\,\sin \th]\,=\,\frac{\omg_c R_{0}^2}{ R_{0}^2 +
r^2 \sin^2 \theta}
\ee
which corresponds to the Newtonian limit of $\omg$ consistent 
with Eq. (\ref{e:lawdif}).
In this limiting case, we now see that $R_0$ is the value of the radius
at which, in the equatorial plane ($\th\,\df\,\pi/2$), the angular velocity is half
its limiting central value, $\omg_c$. Even if this remark is valid only in the
Newtonian case, it is nicely verified for relativistic stars
as well. For all four models we have taken $R_0^2 = 50$ km$^2$, while
$\omg_c$ takes the values of 4500, 3500, 2600, and 1300 rad/s, from
top to bottom. The agreement between the simple law (\ref{omelaw}) and
the results from simulations is quite acceptable. 
In all the models the spatial
scale of variation of the angular velocity ($R_0$) is the same because
the density profiles were very similar (see Figure \ref{fig:modin}, bottom). Varying
the EOS might lead to a different density distribution, in such a way
that one should expect smaller $R_0$ for softer EOSs (more compact
PNS), and larger $R_0$ for stiffer EOSs.  
We must remark again that our assumption of having stationary axisymmetric solutions
implied that the function $F$ on the r.h.s. of the equation of stationary motion
(\ref{e:stmo}) is only a function of the angular velocity. In the Newtonian case,
this is equivalent to say that the angular velocity depends only on the distance to
the axis ($r \sin\th$). In the relativistic case, there are small 
corrections but the overall
distribution of angular velocity is nearly cylindrical, as can be seen in Fig. 
\ref{fig:ome1}. 

In reality, immediately after core collapse and bounce, there is no
reason to expect that the distribution of angular velocity corresponds to that of
stationary equilibrium, and it may take several rotation periods for the star to
relax to some stationary solution. 
A more detailed analysis of
all these issues is out of the scope of this paper, and requires a
parametric study of rotating core collapse simulations. However, for
our purpose, which is just to understand in which range of parameters
one should move when using Eq. (\ref{e:lawdif}), the conclusion is
that a reasonable realistic model with differential rotation should be
consistent with $R_0$ of the order of 10 km.  Considering that there
might be a number of processes that could lead to rigid rotation on a
dynamical timescale, such as shear or magneto-rotational
instabilities, we will study both cases, with substantial
differential rotation and with rigid rotation.  Notice that
previous works (Goussard et al. \cite{G98}) on
differentially rotating PNSs have focused on the parameter space
region corresponding to \mbox{$R_0 \approx 1$ km}, which seems too low
according to our results.

A final remark concerns the geometry of the PNSs.
Although for progenitors with moderate angular momentum,
conservation of angular momentum leads to a characteristic profile of angular
velocity that decreases with increasing distance to the rotation axis, in 
the case of very fast rotation, or unrealistic large differential rotation,
the usual oblate shape of the PNS can become toroidal, with a maximum in density off axis. 
In this case the profile of the angular velocity can exhibit also a 
maximum at some distance from the axis.
We will discuss below for which models a toroidal shape is found, but 
unfortunately our numerical algorithm does not allow us to compute 
equilibrium
configurations with zero density at the origin. 

\section{Results} \label{sec:resnum}

We have considered a number of thermodynamical profiles 
corresponding to snapshots of the early evolution of PNSs, and we have 
calculated the rotational properties in two different cases: rigidly rotating and
differentially rotating PNSs. The left part of Table \ref{tab:m16} shows some properties
(central baryonic density, gravitational mass, and radius)
of the non rotating models with a baryonic mass of \mbox{1.6 $M_\odot$}.
The models are labeled according to the evolutionary time (in s after birth) 
in the non-rotating case.  Notice that for rotating PNSs the 
{\it evolution time} labeling the models is only an indication, since in a 
self-consistent simulation one would expect faster evolution. The reason of
this faster evolution is that the lower
densities, and consistently lower temperatures because the entropy
is conserved,  result in larger neutrino mean free paths. Therefore the neutrino
transport would be more effective.

In order to investigate the influence
of the low density ($<10^{11}$ g/cm$^3$) EOS, we have also substituted the zero 
temperature BPS equation of state (Baym et al. \cite{bps}) 
by the LS (Lattimer \& Swesty \cite{LS91}) EOS in some models at early times. The results
for the models with the LS EOS are shown between parenthesis in the first four
rows of Table \ref{tab:m16}. At later times, after the mantle loses most
of its thermal and lepton content and contracts, the influence of the low density 
EOS becomes less important and both EOSs give essentially the same result, because the
amount of mass and the thickness of the low density layer is rather small.
The left hand side of Table \ref{tab:m12} shows the equivalent information but for a 
non rotating NS with a total baryonic mass of \mbox{1.2 $M_\odot$}.

\subsection{Rigidly rotating PNSs} \label{sec:rigrot}

In the right hand side of Tables \ref{tab:m16} and \ref{tab:m12} 
we summarize the properties of maximally rotating PNSs following 
the same thermodynamical sequence as in the non-rotating case. 
Focusing at early times ($t<0.5$ s)  we see that the central densities of 
non-rotating and maximally rotating
PNSs are not terribly different. The reason is that these objects are only slightly
bounded and they cannot rotate very fast, reaching the mass shedding limit at
angular velocities of hundreds of Hz. Therefore, the rotational energy is not large
and the models are not very different from the non-rotating case. This can also be seen
in the values of the quantity $|T/W|$, which is about 0.03--0.05 for the BPS EOS and
0.06--0.07 when using the LS EOS at low density. The differences between these
two EOSs come from the fact that the BPS is at zero temperature while the LS 
is at finite temperature. Due to the more extended mantle,
the total angular momentum is systematically large, of about 30$\%$, for the models
with finite temperature at $\rho<10^{11}$ g/cm$^3$.

After about 0.5-1.0 seconds, the mantle has contracted and lost its thermal and
lepton content, so that the PNS becomes more compact and bound. 
This can be deduced by looking at the central density,
which exhibits a sudden increase at the end of this first epoch. For the non-rotating
models it happens between 0.5 and 1.0 seconds, while for the maximally rotating models
it seems to be delayed until some time between 1.0 and 2.0 seconds, due to the role
of the centrifugal force. For these more compact configurations, the limiting angular
velocities are much larger (3000-6000 rad/s), and their central densities are considerably
lower than those of the non-rotating PNSs.
The ratio between polar and equatorial radii of maximally rotating PNSs 
turns out to be rather insensitive to the thermodynamics, being about 0.65 for
early models and decreasing to 0.55 for late configurations. With respect to the
$|T/W|$ ratio, its maximum value is about 0.1 at the end of the evolution, 
still
below the threshold to undergo dynamical instabilities.
As a rule of thumb, we found that the maximum angular velocity can be
estimated with good accuracy with the simple fit \mbox{$\Omega_K \approx 0.58\,\sqrt{GM/R^3}$}, valid for both masses (1.6 and 1.2 $M_\odot$) and all 
thermodynamical profiles studied. Despite our models are not close to the
maximum mass for a given EOS, this result is in agreement with the empirical formula
derived by Haensel \& Zdunik (\cite{HZ89}) relating the maximum rotation frequency 
to the mass and radius of the maximum mass static configuration. It seems
that this empirical relation is valid in general.

\begin{figure}
\psfig{figure=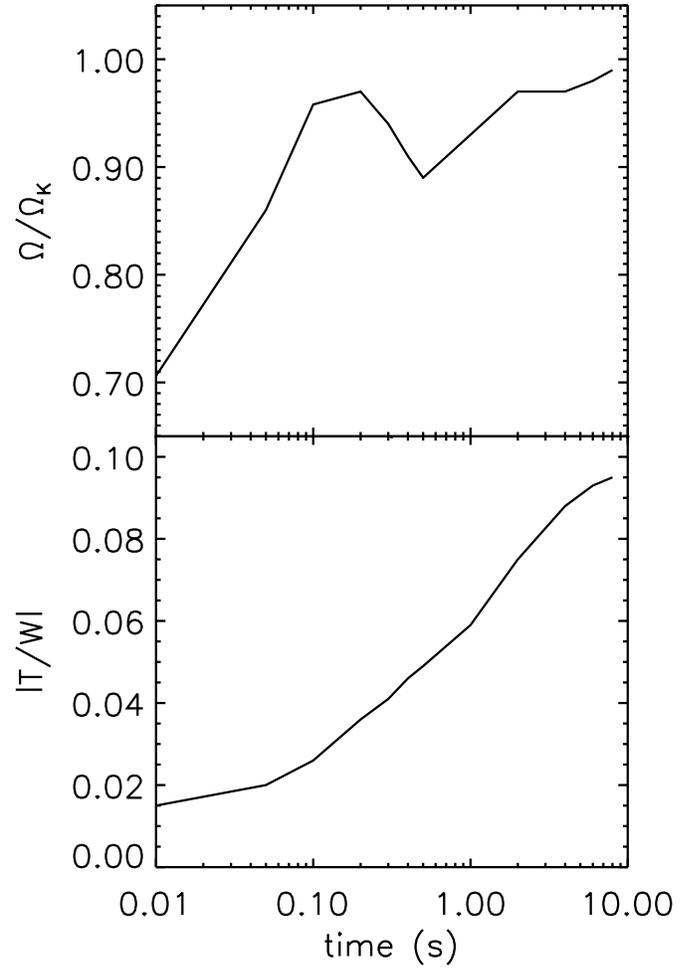}
\caption{Temporal evolution of the angular velocity $\Omega/\Omega_K$ and
the rotation parameter $|T/W|$ for a 
sequence of rigidly rotating PNSs with
constant angular momentum ($1.5 G M_{\sun}^2 /c$)
and a fixed baryonic mass of \mbox{$M_\mathrm{B} = 1.6\,M_{\sun}$}.
}
\label{fig:seqj}
\end{figure}

The evolution of the angular momentum deserves a separate discussion. The first 
result is that including thermal effects at low density (LS) allows for significantly
larger total angular momentum. However, this discrepancy disappears 
rapidly, and the maximum angular momentum decreases 
quite fast and monotonously with time, which has evolutionary implications.
Whether or not neutrino transport can redistribute and take away a substantial
fraction of the initial angular momentum is unclear and needs of multidimensional
transport simulations to be clarified. In general, it is assumed that the angular
momentum is approximately constant during the Kelvin-Helmholtz epoch, 
and it only varies on timescales larger than the diffusion
timescale. If this was the case, PNSs born with maximal angular momentum cannot radiate
it via neutrino diffusion with the same efficiency as the binding energy, and
this may result in  significant mass loss after reaching the mass shedding limit. 
Other possibility is that PNSs are not
born maximally rotating, the angular momentum is (almost) conserved in a timescale
of seconds, and the star speeds up as it contracts and evolves thermodynamically.
We have simulated this scenario by taking a sequence of configurations with 
constant angular momentum \mbox{($J=1.5~ G M_\odot^2/c$)}, always with fixed baryonic mass. 
Results are shown in Figure \ref{fig:seqj}. It can be seen how a model that initially had 
a $|T/W|$ ratio
of only 0.02 speeds up in a timescale of a second and then slowly approaches
the mass shedding limit in a few seconds. 
We expect that fast (but not maximally) rotating PNSs might reach
the threshold of some types of instabilities (CFS, bar--mode, ...) on a diffusion timescale,
say within 10 s after birth. It will depend on the efficiency of neutrinos to
carry away angular momentum, which is currently under investigation.

The above discussion was focused on results for the {\it canonical} NS with a
baryonic mass of 1.6 $M_\odot$. For completeness we also give results for a 
model with a lower baryonic mass of 1.2 $M_\odot$ in Table \ref{tab:m12}.
The results are qualitatively similar to those of Table 1, with only some
quantitative differences. Accordingly to the lower mass, the maximum angular
velocities and $|T/W|$ ratio are also lower.

\begin{table*}
\centering
\caption[]{Properties of non-rotating PNSs and rigidly 
rotating PNSs at the limiting frequency,
for a fixed baryonic mass \mbox{$M_\mathrm{B} = 1.6\,M_{\sun}$}.
The models are labeled by the evolutionary time of non-rotating
PNSs at which the thermodynamical profile was calculated.
The entries in the table are: 
central baryon number density, $n_\mathrm{c}$;
gravitational mass, $M_\mathrm{G}$; 
circumferential equatorial radius, $R_{eq}$;
Kepler frequency, $\Omega_\mathrm{K}$;
angular momentum, $J$; axis ratio $r$ (polar to equatorial)
and the rotation parameter $|T/W|$ (rotational energy on
absolute value of gravitational energy). For 4 time steps at early times
we give results using the LS EOS at low densities for comparison (first four
rows, between parenthesis).}
\label{tab:m16}
\begin{tabular}{ l | c c c | c c c c c c c }
  \hline
  \hline
   & & & & & & & & & \\
Model  & & {$\Omega = 0$} & & & & & {$\Omega = \Omega_\mathrm{K}$} & &  \\
   & & & & & & & & & \\
   \cline{2-11}
   & & & & & & & & & \\
   time & $n_\mathrm{c}$ & $M_\mathrm{G}$ & $R_{eq}$ & 
          $n_\mathrm{c}$ & $M_\mathrm{G}$ & $R_{eq}$ & 
        $\Omega_\mathrm{K}$ & $J$ & r & $|T/W|$ \\
~~(s)  & [fm$^{-3}$] & [$M_{\sun}$]  & [km] 
    & [fm$^{-3}$] & [$M_{\sun}$]  & [km] &
      [rad.Hz]  & [$G M_{\sun}^2 /c$ ] & & \\
   & & & & & & & & & \\
  \hline
   & & & & & & & & & \\
  (0.1)   & (0.006) & (1.595) &  (74.0) & (0.0051) & (1.594) & (112)  & (394.1) & (2.175) & (0.624) & (0.049) \\
  (0.2)   & (0.014) & (1.590) &  (48.6) & (0.0119) & (1.590) & (72.2) & (765.0) & (2.189) & (0.603) & (0.068) \\
  (0.5)   & (0.040) & (1.587) &  (30.7) & (0.0295) & (1.582) & (45.1) & (1565)  & (2.323) & (0.561) & (0.103) \\
  (1.0)   & (0.336) & (1.547) &  (19.1) & (0.0610) & (1.577) & (35.7) & (2239)  & (2.378) & (0.535) & (0.123) \\
\hline
0.1    &  0.006  &  1.592  &  83.5   &  0.0055  &  1.593  &  125   &   332.7 &  1.598  &  0.644  &  0.029  \\
0.2    &  0.014  &  1.592  &   55.3  &  0.0128  &  1.594  &  78.6  &   630.8 &  1.569  &  0.659  &  0.039  \\
0.4    &  0.030  &  1.587  &   37.0  &  0.0267  &  1.584  &  54.0  &  1174   &  1.728  &  0.613  &  0.059  \\
0.5    &  0.040  &  1.585  &   33.3  &  0.0322  &  1.583  &  48.4  &  1384   &  1.780  &  0.606  &  0.066  \\
1.0    &  0.338  &  1.548  &   20.2  &  0.0751  &  1.578  &  37.1  &  2064   &  1.764  &  0.593  &  0.077  \\
2.0    &  0.373  &  1.531  &   16.5  &  0.2006  &  1.557  &  25.9  &  3470   &  1.627  &  0.587  &  0.086  \\
4.0    &  0.399  &  1.508  &   14.4  &  0.2618  &  1.523  &  21.3  &  4644   &  1.632  &  0.565  &  0.101  \\
6.0    &  0.407  &  1.496  &   13.9  &  0.2891  &  1.509  &  20.4  &  4939   &  1.573  &  0.568  &  0.099  \\
8.0    &  0.410  &  1.489  &   13.7  &  0.3043  &  1.501  &  20.0  &  5078   &  1.541  &  0.569  &  0.099  \\
10.    &  0.411  &  1.483  &   13.6  &  0.3156  &  1.495  &  19.7  &  5154   &  1.523  &  0.569  &  0.099  \\
12.    &  0.413  &  1.478  &   13.5  &  0.3248  &  1.491  &  19.6  &  5200   &  1.509  &  0.570  &  0.098  \\
15.    &  0.415  &  1.473  &   13.43 &  0.3346  &  1.487  &  19.5  &  5233   &  1.494  &  0.570  &  0.098  \\
20.    &  0.422  &  1.466  &   13.36 &  0.3444  &  1.483  &  19.4  &  5264   &  1.484  &  0.570  &  0.097  \\
   & & &  & & & & & & \\
  \hline
  \end{tabular}
\end{table*}

\begin{table*}
\centering
\caption[]{Properties of non-rotating PNSs and rigidly 
rotating PNSs at the limiting frequency,
for a fixed baryonic mass \mbox{$M_\mathrm{B} = 1.2\,M_{\sun}$}.
The entries in the table are the same as in Table 1.
} 
\label{tab:m12}
\begin{tabular}{ l | c c c  | c c c c c c c }
  \hline
  \hline
   & & & & & & & & & \\
Model  & & {$\Omega = 0$} & & & & & {$\Omega = \Omega_\mathrm{K}$} & &  \\
   & & & & & & & & & \\
   \cline{2-11}
   & & & & & & & & & \\
   time & $n_\mathrm{c}$ & $M_\mathrm{G}$ & $R_{eq}$ &
         $n_\mathrm{c}$ & $M_\mathrm{G}$ & $R_{eq}$ &
        $\Omega_\mathrm{K}$ & $J$ & r & $|T/W|$ \\
    (s)  & [fm$^{-3}$] & [$M_{\sun}$] & [km] 
    & [fm$^{-3}$] & [$M_{\sun}$]  & [km] &
      [rad.Hz]  & [$G M_{\sun}^2 /c$ ] & & \\
   & & & & & & & & & \\
  \hline
   & & & & & & & & & \\
    0.1   &  0.0065  &   1.199  &  78.73  &  0.0060   &  1.199  &  117.6   &  314.7  &  0.876  &  0.65  &  0.0241 \\
    0.4   &  0.038   &   1.194  &  32.80  &  0.0302   &  1.191  &   47.91  & 1216  &  1.018  &  0.62  &  0.0569 \\
    0.5   &  0.057   &   1.193  &  29.25  &  0.0390   &  1.191  &   43.05  & 1429  &  1.043  &  0.61  &  0.0631 \\
    0.6   &  0.096   &   1.191  &  26.36  &  0.0495   &  1.190  &   39.71  & 1614  &  1.051  &  0.61  &  0.0673 \\
    0.75  &  0.248   &   1.174  &  20.87  &  0.0673   &  1.187  &   36.40  & 1836  &  1.030  &  0.60  &  0.0689 \\
    1.    &  0.260   &   1.172  &  19.38  &  0.1078   &  1.187  &   32.27  & 2204  &  0.989  &  0.60  &  0.0689 \\
    2.    &  0.286   &   1.156  &  15.58  &  0.1893   &  1.165  &   22.93  & 3577  &  0.980  &  0.59  &  0.0870 \\
    3.    &  0.295   &   1.149  &  14.79  &  0.2136   &  1.154  &   21.62  & 3965  &  0.945  &  0.58  &  0.0880 \\ 
    4.    &  0.298   &   1.144  &  14.45  &  0.2271   &  1.149  &   20.97  & 4138  &  0.926  &  0.58  &  0.0877 \\ 
    5.    &  0.299   &   1.140  &  14.25  &  0.2363   &  1.145  &   20.61  & 4237  &  0.912  &  0.58  &  0.0874 \\
    6.    &  0.300   &   1.137  &  14.12  &  0.2439   &  1.142  &   20.39  & 4300  &  0.903  &  0.58  &  0.0871 \\
    7.    &  0.301   &   1.134  &  14.03  &  0.2498   &  1.140  &   20.24  & 4340  &  0.897  &  0.58  &  0.0868 \\
    8.    &  0.302   &   1.132  &  13.96  &  0.2553   &  1.138  &   20.14  & 4370  &  0.891  &  0.58  &  0.0866 \\
   10.    &  0.306   &   1.129  &  13.88  &  0.2638   &  1.135  &   20.03  & 4400  &  0.882  &  0.58  &  0.0860 \\
   12.    &  0.311   &   1.126  &  13.82  &  0.2698   &  1.134  &   19.96  & 4420  &  0.878  &  0.58  &  0.0859 \\
   & & & & & & & & & \\
  \hline
  \end{tabular}
\end{table*}

\subsection{Differentially rotating PNSs}  \label{sec:difrot}

As discussed in the previous section we will restrict to realistic
values of $R_0$ in the range from 10 km (significant differential
rotation) to 50 km (almost rigid rotation).  A different scenario
would be a hot NS resulting from the merger of two neutron stars. In
this case, the distribution of angular momentum is more complex and
depends not only on the initial total angular momentum, but also on a
number of different parameters as the mass ratio or the relative
orientation of the spins of the two stars. Lower values of $R_0$
might be reached in this latter scenario, but above all, the rotation
profile would probably be very different from Eq. (\ref{e:lawdif}),
which is in agreement with a stationary configuration, the case of a
merging being far from a stationary situation.

\begin{figure}
\psfig{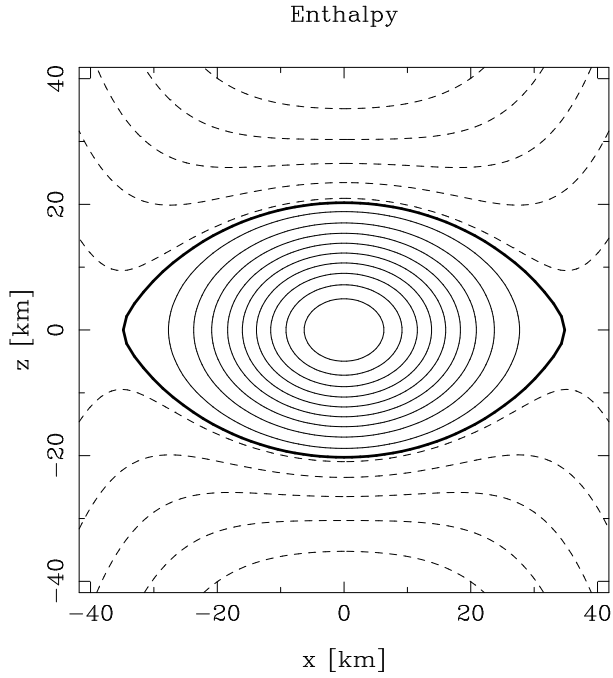}
\caption{Iso-enthalpy lines 
of a differentially rotating star with \mbox{$R_0=100$ km}, for $t=1$ s, 
and rotating with $\Omega = 354$ rad/s. ($\Omega/\Omega_K=0.995$). For
low differential rotation the structure of the star still looks 
{\it normal}.
\label{fig:enth1}
}
\end{figure}

\begin{figure}
\psfig{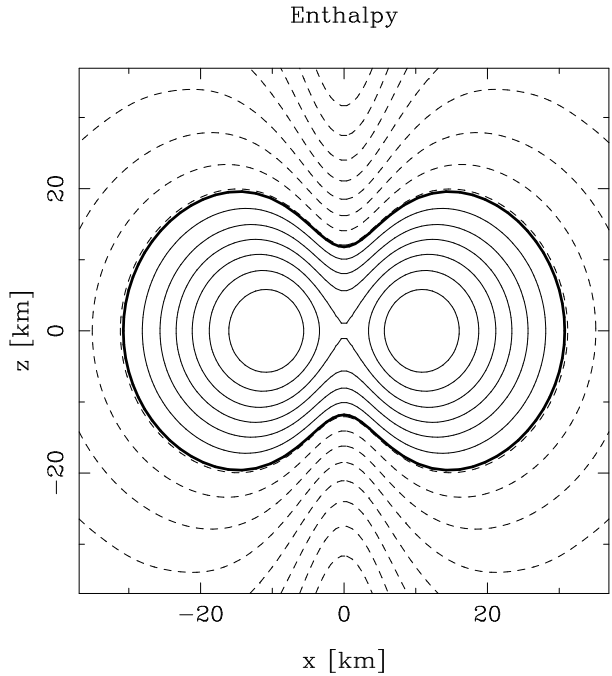}
\caption{Iso-enthalpy lines 
of a differentially rotating star with \mbox{$R_0=10$ km}, for $t=1$ s, 
and rotating with $\Omega = 1570$ rad/s. The {\it toroidal} shape is
clearly visible for all models differentially rotating with $R_0$ smaller
than a certain threshold.
}
\label{fig:enth2}
\end{figure}

As before, the baryonic mass of the star is conserved and fixed to 1.6 $M_\odot$
for all models. Within the range of parameters mentioned above,
we have calculated a large number of models of differentially rotating PNSs
with either a fixed $R_0$ of 10, 20, 50, and 100 km or varying $R_0$ according
to the circumferential equatorial radius of the star (\mbox{$R_0\,=\,0.5\,R_{eq}$}). 
In Figures \ref{fig:enth1} and \ref{fig:enth2} we show iso-enthalpy surfaces
for two representative models at an evolutionary time of \mbox{$t=1$ s}.
Despite both configurations have similar equatorial radii ($\approx 30$ km),
they are qualitatively very different.
In Figure \ref{fig:enth1} the parameter $R_0=100$ km, {\it i.e.}, the 
model has
a very low degree of differential rotation. Even very close to the mass shedding limit
(0.995 of the maximum angular velocity) the star shows a regular oblate profile.
On the contrary, the model in  Figure \ref{fig:enth2} corresponds to the case
of significant differential rotation ($R_0=10$ km). With strong differential
rotation we have found that above some critical
angular velocity (950 rad/s, for this model) the shape of the star becomes 
toroidal, with a maximum density off axis.

\begin{figure}
\psfig{figure=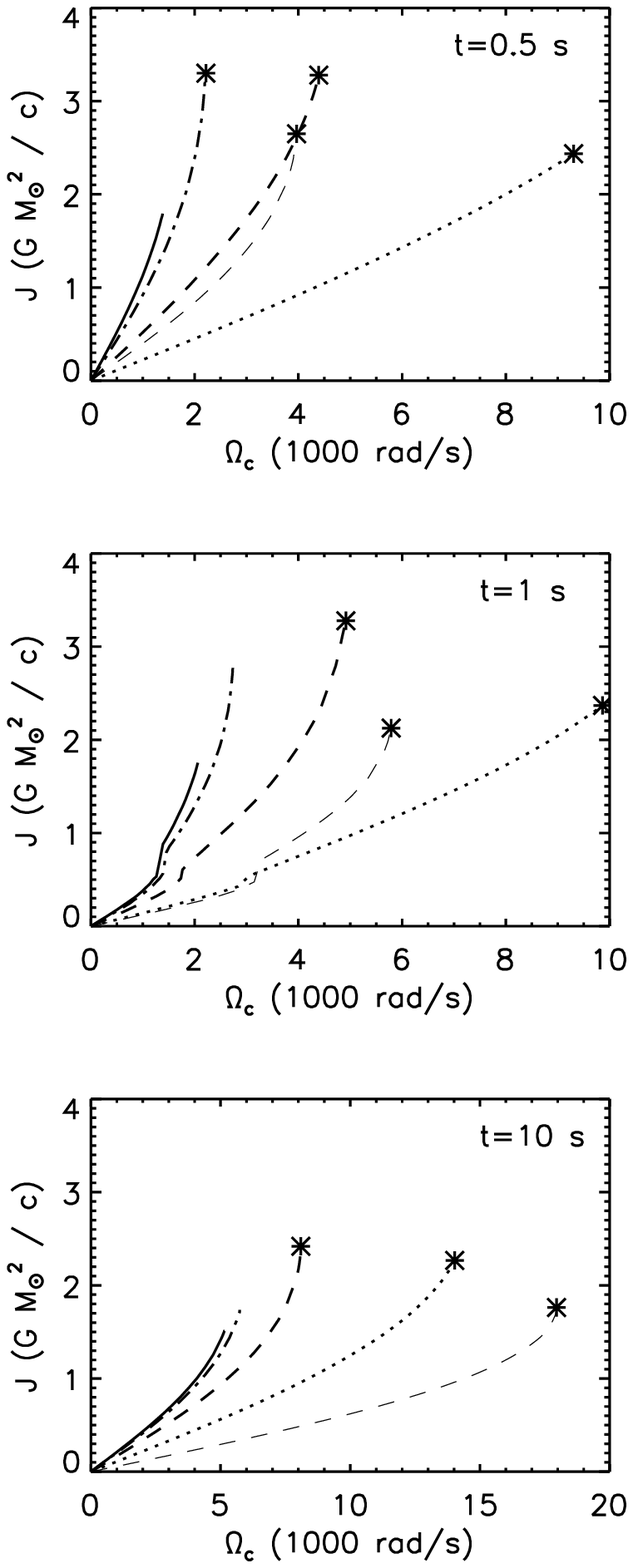}
\caption{ Total angular momentum as a function of the central
angular velocity for PNSs with fixed baryonic mass of 1.6 $M_\odot$ at
evolutionary times of 0.5 (top), 1.0 (middle) and 10 s (bottom),
and for different values of $R_0$:
$R_0=\infty$ (solid), 50 km (dash-dot), 
20 km (thick dashed), $R_{eq}/2$ (thin dashed), 10 km (dots).
}
\label{figj}
\end{figure}

\begin{figure}
\psfig{figure=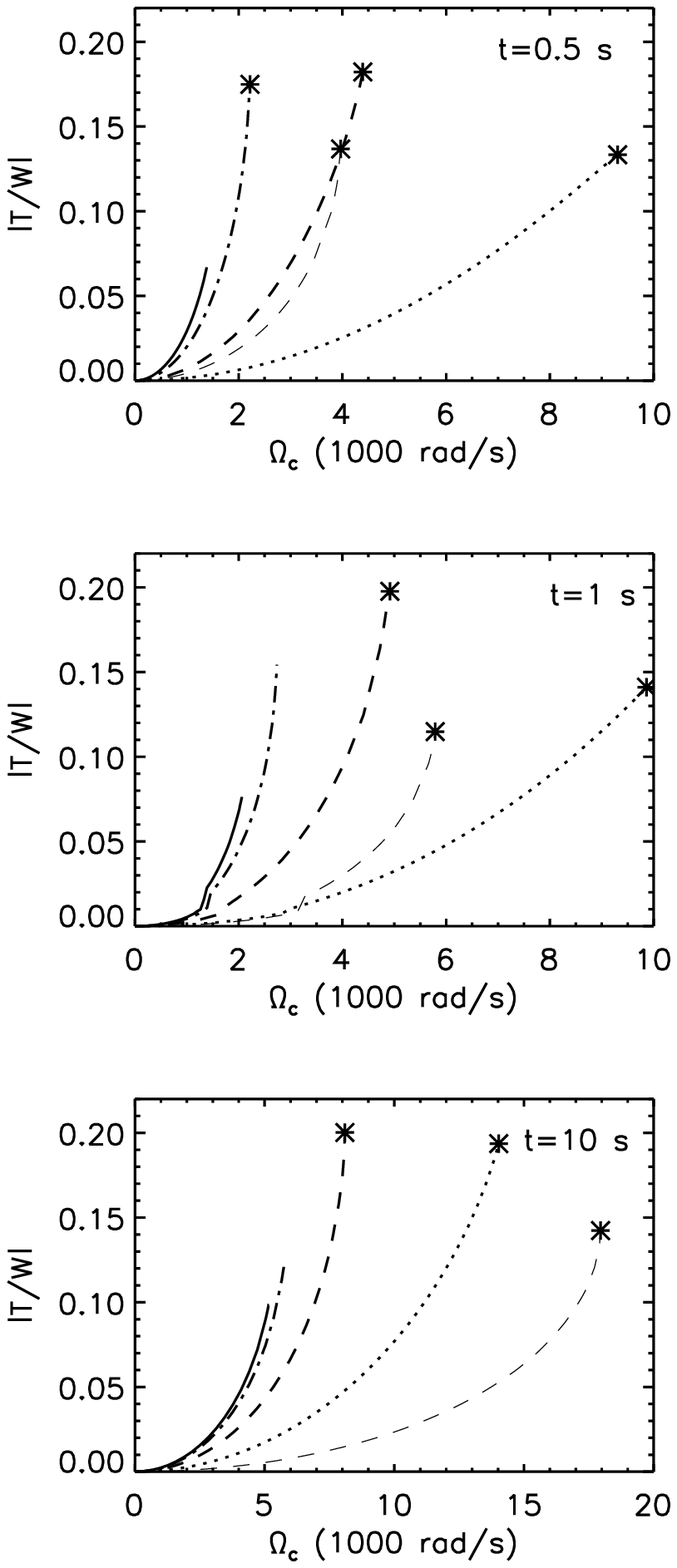}
\caption{ Rotation parameter, $|T/W|$, as a function of the central
angular velocity for PNS with fixed baryonic mass of 1.6 $M_\odot$ at
evolutionary times of 0.5 (top), 1.0 (middle) and 10 s (bottom),
and for different values of $R_0$:
$R_0=\infty$ (solid), 50 km (dash-dot), 
20 km (thick dashed), $R_{eq}/2$ (thin dashed), 10 km (dots).
}
\label{figtw}
\end{figure}

The most relevant results are presented in
a compact form in Figures  \ref{figj} and  \ref{figtw}, where we
show the total angular momentum and the rotation parameter $|T/W|$
as a function of the central angular velocity $\Omega_c$, for different values
of $R_0$. For comparison, we have also included the results for rigid rotation
(solid curve), together with those for 
$R_0=50$ km (dash-dotted line), $R_0=20$ km (thick dashes),
$R_0=R_{eq}/2$ (thin dashes) and $R_0=10$ km (dots). 
For $R_0\ge100$ km, the results are
so close to the rigid rotation case that we do not show them in the figures.

The thermodynamical profiles for each panel  
are those corresponding to evolutionary times of the non-rotating models
of $t=0.5$ (top), $1.0$ (middle) and 10 s (bottom panel) after formation. 
From the results of the figure we can see that for $R_0$ of the order
or larger than 50 km the effects of differential rotation are 
unimportant. Only for $R_0 < 20$ km and $R_0=0.5 R_{eq}$ the differences are significant. 
As discussed in \ref{sec:inimod}, results from axisymmetric core collapse seem to indicate
that the scale in which one should expect variations of the angular velocity
is of about 10 km, therefore the last two cases may be looked at as the most
realistic, with the caveat of being constrained to the particular rotation law
we used.

In the model corresponding to $t=1$ s it is visible a sudden increase
of the angular momentum at a critical value of the angular velocity,
different for each value of $R_0$. This is an effect of the phase
transition from bulk matter to nuclei, and it is associated to the point 
when the angular velocity becomes so large that the central density is
below nuclear saturation density ($\rho_0$).  This transition is not
observed at earlier times because matter at low density is still very hot,
and the thermal effects suppress or smooth out the kink. At later times, 
the star is more compact and the central density is always above $\rho_0$,
even close to the mass shedding limit.
For the models with strong differential rotation, we had to stop our calculations
before reaching the mass shedding limit because of numerical limitations associated
to the special toroidal geometry that arises (see Fig. \ref{fig:enth2}). These cases
are marked by an asterisk in Fig. \ref{figj} and  \ref{figtw}. In principle, there
could be configurations with slightly larger angular momentum.

\begin{figure}
\psfig{figure=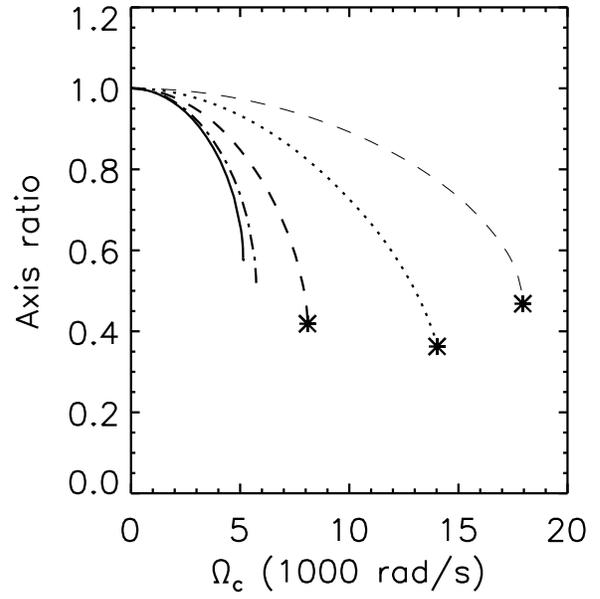}
\caption{ Axis ratio (polar to equatorial)  as a function of the central
angular velocity for a PNS with a baryonic mass of 1.6 $M_\odot$ at
$t=10$ s,
and for different values of $R_0$:
$R_0=\infty$ (solid), 50 km (dash-dot), 20 km (thick dashed), $R_{eq}/2$
(thin dashed), 10 km (dots).
}
\label{figaxis}
\end{figure}

An expected result is that the central angular velocity can be up to a factor 5 to 10
larger in the case of differential rotation, but the maximum angular momentum is only
about a fifty percent larger than for rigid rotation. More important is the variation
in the maximum of  $|T/W|$. While for rigidly rotating stars it is about 0.1, for
PNSs with a large degree of differential rotation it can be as high as 0.2. 
The fact that differential rotation allows for the existence of equilibrium models
at large  $|T/W|$ have also been found by Baumgarte et al. (\cite{Bau00})
and Yoshida et al.
(\cite{YRKE}) for polytropes in the relativistic Cowling approximation using the
same rotation law. A maximum value of $|T/W|=0.2$ is not enough to reach the 
dynamical instability threshold ($\approx 0.27$), but it is 
sufficient to allow either the secular, gravitational wave driven, instability 
(Chandrasekhar \cite{cha}; Friedman \& Schutz \cite{FS})  that happens at $0.1-0.15$
or the recently proposed low $|T/W|$ dynamical instabilities
(Shibata et al. \cite{Shi02,Shi03}; Watts et al. \cite{WAJ}).
Notice that despite our calculations cannot be carried up to the mass shedding
limit for some models, we are not far from it and the numbers are close to the
maximum value. This can be seen in Figure \ref{figaxis}, when we show the polar
to equatorial axis ratio as a function of $\Omega_c$. For rigid rotation, or moderate
differential rotation the calculations
end at the Keplerian limit, while for strong differential rotation the calculation
ends because of numerical problems (marked with asterisks). The slope of the curves is 
becoming more and more vertical, and the axis ratio is as low as 0.4, smaller than the
minimum in the rigid case (0.6). This is an indication that the shape of the
star is changing fast and only slightly larger angular
velocities can be reached before the star starts loosing mass. 

\begin{figure}
\psfig{figure=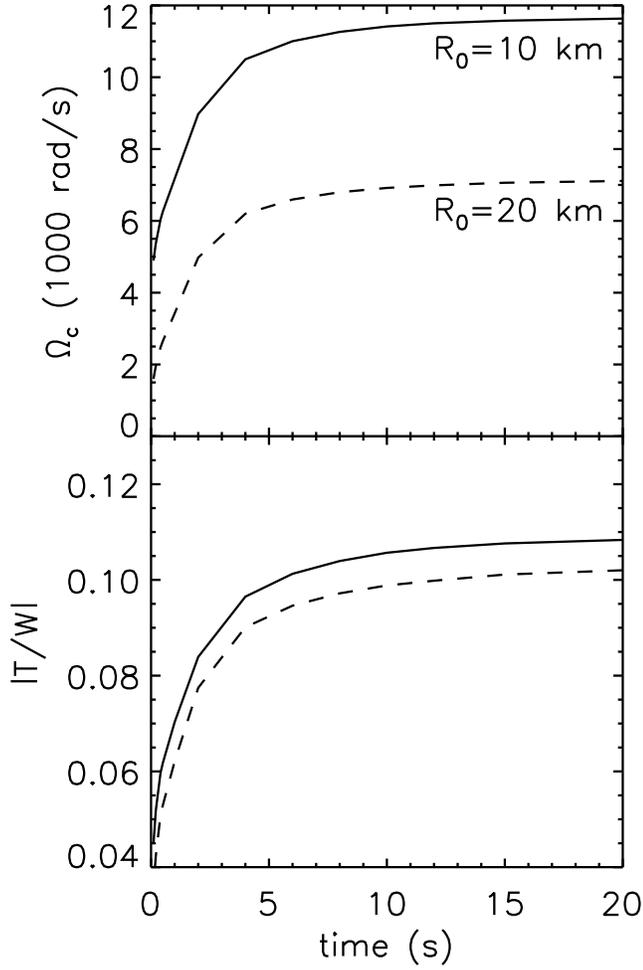}
\caption{Temporal evolution of the central angular velocity $\Omega_c$
and of the rotation parameter $|T/W|$ for two sequences of
differentially rotating PNSs with $R_0=10$ km (solid line) and
$R_0=20$ km (dashed line). In both cases we fix the total angular
momentum to $1.5\,G M_{\sun}^2 /c$ and the baryonic mass to
\mbox{$M_\mathrm{B} = 1.6\,M_{\sun}$}.  }
\label{fig:seqjdif}
\end{figure}

The next question one can pose is whether or not dynamical instabilities
arise in a real case.
As for rigidly rotating stars, we now impose that the angular momentum is
fixed along the evolution, as well as the baryonic mass, and study how
the different variables evolve under such constraints. Figure \ref{fig:seqjdif}
shows the temporal evolution of the central
angular velocity (top) and of the $|T/W|$ ratio (bottom)
of a PNS differentially rotating with $R_0=10$ km and a fixed angular momentum of
\mbox{$J=1.5~ G M_\odot^2/c$}. 
As in the case of rigid rotation, $|T/W|$ increases as the star contracts 
and looses its binding energy in a neutrino diffusion timescale of about 10 s. Thus,
if there are no significant losses of angular momentum, the various types 
of dynamical 
instabilities associated to critical values of the rotation parameter may 
arise
not at the very beginning, but several seconds after formation.

\begin{figure}
\psfig{figure=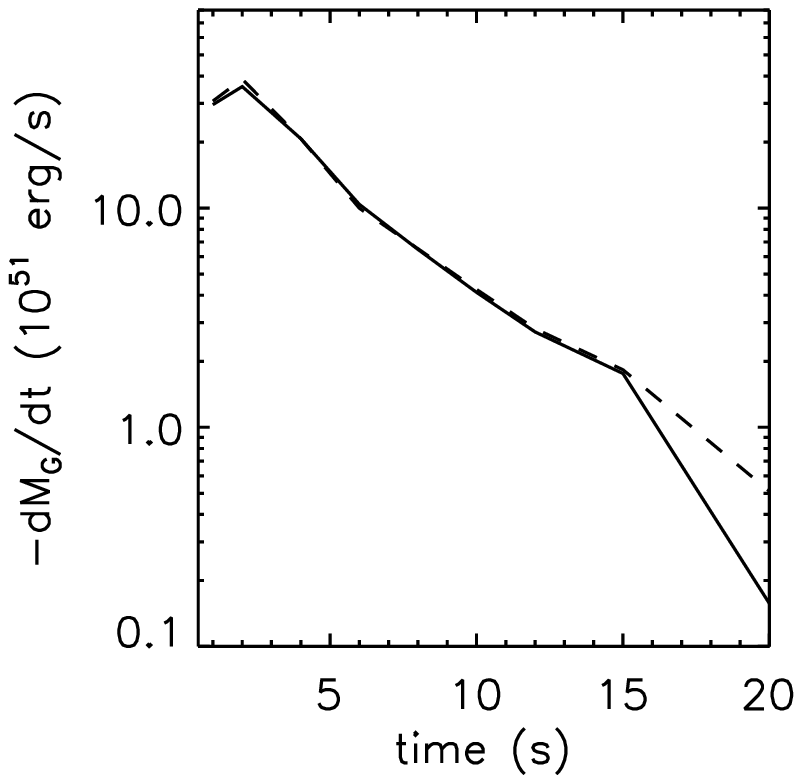}
\caption{Estimated neutrino luminosity ($-dM_G/dt$) as a function of
time, consistent with the evolution at fixed baryonic mass of
\mbox{$M_\mathrm{B} = 1.6\,M_{\sun}$} and angular momentum of \mbox{$1.5\,G
M_{\sun}^2 /c$}. The two models are differentially rotating with
\mbox{$R_0=10$ km} (solid line) and \mbox{$R_0=20$ km} (dashed line).
The order of magnitude and the exponential decay are similar to the
luminosities obtained in simulations with neutrino transport for
non-rotating stars.  }
\label{fig:lum}
\end{figure}

As a consistency check of our assumptions and many simplifications, one can
make a simple estimate of the neutrino luminosity considering that
the variation in the gravitational mass of the PNS is considered to be equal to the
luminosity,{\it  i.e.}, $L_{\nu} = - \frac{d M_G}{dt}$. The result is 
shown in
Figure \ref{fig:lum}. Both the order of magnitude and the exponential decay
of the luminosity in a timescale of 10 seconds are consistent with the
results of detailed simulations with neutrino transport in spherical symmetry
(Pons et al. \cite{pns99}). 
It is interesting that the difference between the two models studied with
$R_0=10$ km (solid) and $R_0=20$ km (dashes) is only visible after 10 seconds
of evolution, when the stars become more compact.
Therefore, imposing conservation of angular
momentum and baryonic mass, and building a sequence keeping fixed $R_0$ is not
violating any basic physical law, such as conservation of energy, and can be 
taken as a qualitative approach to the real case. 

\section{Conclusions} \label{sec:conc}

We have approached the problem of the evolution of rotating protoneutron stars
by constructing evolutionary sequences of axisymmetric stationary configurations
in General Relativity. The thermodynamical structure and evolution 
have been taken and extrapolated
from simulations in spherical symmetry that included neutrino transport. Although
this is a crude simplification, it already gives an interesting insight about
how the different relevant quantities can evolve as the rotating PNS loses
its lepton content and its excess binding energy, and contracts. Moreover,
we have found that the luminosity estimates are not terribly different from
what one expects.

A special effort has been made to understand in which space
of parameters we should be in a realistic case. The biggest uncertainty concerns
the rotation law that PNS have at birth. By analyzing results from simplified
core collapse simulations, it seems that a typical scale for variations of
the angular velocity is about 10 km, and that conservation of angular momentum
during the collapse of a stellar core (initially rigidly rotating) does not seem
to allow for angular velocities varying in a length--scale shorter than a few km.
Less is known about the angular distribution, except that the most recent simulations
show the presence of important meridional currents and some turbulent motion.
For simplicity, we restricted ourselves to the stationary case. Stationarity
implies a quasi-cylindrical distribution (with deviations due to relativistic corrections)
of the angular velocity. This stage only can be reached after several
dynamical and rotation periods, after the PNS had time to relax.
Therefore one must be aware that the first 0.5 s are probably far from stationarity,
but after that evolution proceeds in a quasi-stationary way, except for
low velocity convective motions. From our study of quasi-stationary sequences we can 
draw a few interesting qualitative results.

i) For rigidly rotating stars, the mass shedding limit is well approximated by
the simple law \mbox{$\Omega_K \approx 0.58 \sqrt{GM/R^3}$}, despite complications
in the EOSs. This empirical formula is valid for stars of different masses within
a 5\% accuracy for the EOSs we considered. We believe that this result is general
and would be valid for other EOSs or stars with different masses.

ii) For each instant in the evolution, stars with strong differential rotation
can have 5 to 10 times larger central angular velocities, and accommodate about a fifty
percent more angular momentum. The maximum specific angular momentum $J/M$ varies between
$(1-2)~ G M_\odot/c \approx (0.5-1) \times 10^{16}$ cm$^2$/s, depending on 
the degree of differential rotation.

iii) When $R_0 \le 20$ km there is a critical value of the angular velocity 
above which the shape of the star becomes toroidal, with a maximum density
off axis. We found that this configurations are stationary solutions of the 
hydrodynamics equations, 
but it is unclear under which circumstances this geometry is stable against perturbations. 
This issue deserves a separate study.

iv) The maximum value of $|T/W|$ obtained in the case of differential 
rotation is about 0.2, while for rigid rotation this is $\approx 0.11$.
Thus differentially rotating PNSs might undergo the CFS instability,
which arises at  $\approx 0.14$
and, in any case, the recently discussed low $|T/W|$ instability
(Shibata et al. \cite{Shi02,Shi03}; Watts et al. \cite{WAJ})
is plausible to happen.

v) More interestingly, we found several situations in which, even if the
initial model is not close enough to the critical value of $|T/W|$, as the star
contracts in a neutrino diffusion timescale of 5-10 s, it speeds up
entering the window of instability. An observational evidence of this effect
could be a temporal shift of a few seconds between the neutrino luminosity
peak and a gravitational wave burst in the event of a galactic Supernova.
Ultimately, this  depends on the initial amount of angular momentum, which
is approximately equal to the angular momentum of the iron core of the
progenitor. Recent stellar evolution calculations suggest that the specific
angular momentum of the inner 1.7 $M_\odot$ of a 15 $M_\odot$ star can be
as high as $3\times10^{16}$ cm$^2$/s if magnetic braking is neglected,
or $10^{15}$ cm$^2$/s if magnetic torques are included in the evolution
(Heger et al. \cite{Heg03}). This corresponds to $J/M$ in the range
$J/M=(0.2-6)~ G M_\odot/c$. 
If the angular momentum happens to be in the upper region ($J/M>2$), 
centrifugal forces would stop the 
collapse before the PNS is formed. Intermediate values ($J/M=1$)
may result in the formation of a rapidly rotating PNS that enters
the instability region several seconds after birth. If magnetic braking is
very effective during the evolution of a massive star, $J/M < 0.5$ and 
the PNS will be formed after collapse without reaching extreme values of the angular
velocities and $|T/W|$.

The next natural step to improve this work is to include the possible mechanisms
to transport angular momentum between the different layers of the star, that may
involve neutrino transport, turbulent transport, magnetic fields, neutrino viscosity, 
convective motion and/or
angular momentum losses by gravitational wave emission.
Unless the star is born with almost maximal angular velocities, since the different 
instabilities can arise in a timescale of seconds some of all of these dissipative
mechanisms can modify our current vision of PNS evolution.


\section*{Acknowledgments}
We are grateful to H. Dimmelmeier for providing us with rotation
profiles that we have used to constrain the parameters of the rotation
law. We thank L. Rezzolla, J. Novak, S. Yoshida and D. Gondek-Rosinska
for useful discussions and comments. This work has been supported by
the EU Programme `Improving the Human Research Potential and the
Socio-Economic Knowledge Base' (Research Training Network Contract
HPRN-CT-2000-00137) and the Spanish Ministerio de Ciencia y Tecnologia
grant AYA 2001-3490-C02. JAP is supported by a {\it Ram\'on y Cajal}
contract from the Spanish MCyT and LV benefited from the Jumelage
PAN-CNRS Astronomy France-Poland.

\begin{thebibliography}{}

\bibitem[2000]{Bau00}
Baumgarte, T.W., Shapiro, S.L., \& Shibata, M. 2000, ApJ, 528, L29

\bibitem[1971]{bps}
Baym, G., Pethick, C.J., \& Sutherland, J. 1971, ApJ, 170, 299 

\bibitem[1993]{BGSM}
Bonazzola, S., Gourgoulhon, E., Salgado M., \& Marck J.-A.
1993, A\&A 278, 421

\bibitem[2003]{BRU}
Bonanno, A., Rezzolla, L., \& Urpin, V. 2003,  A\&A, 410, L33 

\bibitem[1986]{bl86}
Burrows A., \& Lattimer J. M. 1986, ApJ, 307, 178

\bibitem[1969]{c69}
Carter B. 1969, J. of Math. Phys., 10, 70

\bibitem[1970]{cha}
Chandrasekhar, S. 1970, Phys. Rev. Lett., 24, 611

\bibitem[2002]{DFM}
Dimmelmeier, H., Font, J.A., \& M\"uller, E. 2002, A\&A, 393, 523

\bibitem[1978]{FS}
Friedman, J.L., \& Schutz, B.F. 1978, ApJ, 221, 937

\bibitem[1999]{GHLPBM}
Gourgoulhon, E., Haensel, P., Livine, R., et al. 1999, A\&A, 349, 851 

\bibitem[1997]{G97}
Goussard, J.O., Haensel, P., \& Zdunik, J.L. 1997, A\&A, 321, 822
\bibitem[1998]{G98}
Goussard, J.O., Haensel, P., \& Zdunik, J.L. 1998, A\&A, 330, 1005

\bibitem[1989]{HZ89}
Haensel, P., \& Zdunik, J.L. 1989, Nature, 340, 617

\bibitem[2000]{Heg00}
Heger, A., Langer, N., \& Woosley, S.E. 2000, ApJ, 528, 368

\bibitem[2003]{Heg03}
Heger, A., Woosley, S.E., Langer, N., \& Spruit H. 2003, 
{\bf \rm astro-ph/0301374},
to appear in {\it Stellar Collapse} (Astrophysics and Space Science) edited by 
C.L. Fryer

\bibitem[1977]{Kaz77}
Kazana, D. 1977, Nature, 267, 501

\bibitem[1986]{kj95}
Keil, W., \& Janka H-Th. 1995, ApJ, 295, 146

\bibitem[1996]{KJM96}
Keil, W., Janka, H-Th., \& M\"uller, E. 1996, ApJ, 473, L111

\bibitem[1989]{keh89}
Komatsu, H., Eriguchi, Y., \& Hachisu, I. 1989, MNRAS, 239, 153

\bibitem[1991]{LS91}
Lattimer, J.M., \& Swesty, F.D. 1991, Nucl. Phys., A535, 331

\bibitem[2002]{MPU}
Miralles, J.A., Pons, J.A., \& Urpin, V. 2002, ApJ, 574, 356

\bibitem[2003]{Ewa03}
M\"uller, E., Rampp, M., Buras, R., Janka, H-Th., \&  Shoemaker, D.H. 2003, 
{\bf \rm astro-ph/0309833}, submitted to ApJ

\bibitem[1990]{MS90}
Myers, W.D., \& Swiatecky, W.J. 1990, Ann. Phys.(NY), 204, 401

\bibitem[1998]{NSGE}
Nozawa, T., Stergioulas, N., Gourgoulhon, E., \& Eriguchi, Y. 1998,
A\&A Suppl. Ser., 132, 431

\bibitem[1999]{pns99}
Pons, J.A., Reddy, S., Prakash, M., Lattimer, J.M., \& Miralles, J.A. 1999, 
ApJ, 513, 780
\bibitem[2000]{pns00}
Pons, J.A., Miralles, J.A., Prakash, M., \& Lattimer, J.M. 2000, ApJ, 553, 382
\bibitem[2001]{pns01}
Pons, J.A., Steiner, A.W., Prakash, M., \& Lattimer, J.M. 2001, 
Phys. Rev. Lett., 86, 5223

\bibitem[1997]{Pra97}
Prakash M. et. al, 1997, Phys. Rep. 280, 1

\bibitem[1992]{rom92}
Romero J.V., Alonso, J.D., Ib\'a\~{n}ez, J.M., Miralles, J.A., \& P\'erez, A. 1992, ApJ, 395, 612

\bibitem[2002]{Shi02}
Shibata, M., Karino, S., \& Eriguchi, Y. 2002, MNRAS, 334, L27

\bibitem[2003]{Shi03}
Shibata, M., Karino, S., \& Eriguchi, Y. 2003, MNRAS, 343, 619

\bibitem[1999]{SSW99}
Strobel, K., Schaab, Ch., \& Weigel, M.K. 1999, A\&A, 350, 497

\bibitem[1999]{sumi99}
Sumiyoshi, K., Ib\'a\~{n}ez, J.M., \& Romero, J.V. 1999, A\&A Suppl. Ser., 134, 39

\bibitem[1996]{S96}
Swesty, F.D., 1996, J. Comp. Phys., 127, 118

\bibitem[2003]{WAJ}
Watts, A.L., Andersson, N., \& Jones, D.I. 2003, {\bf \rm astro-ph/0309554}, submitted to MNRAS

\bibitem[2002]{YRKE}
Yoshida, S., Rezzolla, L., Karino, S., \& Eriguchi, Y. 2002, ApJ, 568, L41

\bibitem[2003]{YH03}
Yuan, Y., \& Heyl, J.S., 2003, astro-ph/0305083

\end{thebibliography}
\end{document}